\documentclass[prb,longbibliography,reprint,superscriptaddress,citeautoscript]{revtex4-1}

\usepackage[applemac]{inputenc}
\usepackage[T1]{fontenc}
\usepackage[american]{babel}
\usepackage[scaled]{helvet}
\usepackage{graphicx,amssymb,bm,mathtools,textcomp,courier,multirow,multirow,bigdelim,color}
\usepackage[normalem]{ulem}	

\newcommand{\op}[2]{\left.{\left| #1\right\rangle\left\langle #2\right|}\right.}

\newcommand{\ket}[1]{\left|\left. #1 \right\rangle\right.}

\newcommand{\figureshortname}{Fig.}
\newcommand{\equationshortname}{Eq.}

\newcommand{\qu}{\mathcal{D}}
\newcommand{\an}{\mathcal{A}}

\newcommand{\meref}[1]{Eqs.~\eqref{#1}}
\newcommand{\eref}[1]{\equationshortname~\eqref{#1}}
\newcommand{\sref}[1]{Sec.~\ref{#1}}
\newcommand{\aref}[1]{Appx.~\ref{#1}}
\newcommand{\cref}[1]{Chapter~\ref{#1}}
\newcommand{\fref}[1]{\figureshortname~\ref{#1}}

\newcommand{\rcite}[1]{Ref.~[\onlinecite{#1}]}
\newcommand{\mrcite}[2]{Refs.~[\onlinecite{#1},\onlinecite{#2}]}

\graphicspath{{Pictures/}}

\begin{document}
\def\sectionautorefname{Sec.}

\title{Fault-Tolerant Quantum Computation for Singlet-Triplet Qubits with Leakage Errors}

\author{Sebastian Mehl}
\email{s.mehl@fz-juelich.de}
\affiliation{Peter Grünberg Institute (PGI-2), Forschungszentrum Jülich, D-52425 Jülich, Germany}
\affiliation{JARA-Institute for Quantum Information, RWTH Aachen University, D-52056 Aachen, Germany}

\author{Hendrik Bluhm}
\affiliation{JARA-Institute for Quantum Information, RWTH Aachen University, D-52056 Aachen, Germany}

\author{David P. DiVincenzo}
\affiliation{Peter Grünberg Institute (PGI-2), Forschungszentrum Jülich, D-52425 Jülich, Germany}
\affiliation{JARA-Institute for Quantum Information, RWTH Aachen University, D-52056 Aachen, Germany}

\date{\today}

\begin{abstract}
We describe and analyze leakage errors of singlet-triplet qubits. Even though leakage errors are a natural problem for spin qubits encoded using quantum dot arrays, they have obtained little attention in previous studies. We describe the realization of leakage correction protocols that can be implemented together with the quantum error correction protocol of the surface code. Furthermore we construct explicit leakage reduction units that need, in the ideal setup, as few as three manipulation steps. Our study shows that leakage errors can be corrected without the need of measurements and at the cost of only a few additional ancilla qubits and gate operations compared to standard quantum error correction codes.
\end{abstract}

\maketitle

\section{Introduction}
Singlet-triplet qubits (STQs) are an excellent candidate for the realization of quantum computation \cite{levy2002,taylor2005,taylor2007}. They are a variety of spin qubit \cite{loss1998}, which is coded on the $s_z=0$ subspace of two electrons that are trapped at a double quantum dot (DQD) \cite{levy2002}. Universal single-qubit control is provided by the exchange interactions between the electrons, when the setup is operated at large magnetic fields with a small, time-independent magnetic field gradient across the DQD \cite{hanson2007-1}.
The magnetic field gradient causes different phase evolutions of $\ket{\uparrow\downarrow}$ and $\ket{\downarrow\uparrow}$, i.e., when the electrons with antiparallel spin configurations are spatially separated at different quantum dots (QDs). We call this operation a phase gate. If the electrons can tunnel between the QDs, then the doubly occupied configuration at a QD can be approached. Because for a doubly occupied QD the Pauli exclusion principle favors the singlet over the triplet configurations, the difference in the phase evolution of the singlet and the spinless triplet can be controlled through the coupling strength of the DQD. Subnanosecond manipulations of the exchange couplings were realized experimentally using electric signals that detune the QD potentials\cite{petta2005}. We call these operations exchange gates.
Universal single-qubit control has been realized for STQs coded using GaAs DQDs \cite{foletti2009,bluhm2010,shulman2014} and Si DQDs \cite{maune2012,wu2014}.

Two-qubit gates  between neighboring DQDs were proposed theoretically using exchange interactions \cite{levy2002}, Coulomb interactions \cite{taylor2005}, or mediated couplings via a cavity \cite{burkard2006,taylor2006-2} or via another QD \cite{mehl2014-1}. Also the first steps towards the experimental realizations of two-qubit gates have been done \cite{weperen2011,shulman2012,frey2012,toida2013}. Furthermore the initialization and the readout of STQs has been successfully achieved using the Pauli spin blockade \cite{petta2005,hanson2007-2,zwanenburg2013}. This paper extends the discussions of fault-tolerant quantum computation \cite{divincenzo2000-2} for STQs, assuming that the initialization, readout, and universal qubit control have high fidelities.

The leakage of quantum information out of the coding subspace is a generic problem for quantum computers. Because gate operations, as for spin qubits \cite{levy2002,divincenzo2000} or superconducting qubits \cite{strauch2003}, use couplings to states that are not part of the qubit subspace, the manipulations of quantum states increase the probability of leakage. Besides optimized gate sequences for qubit manipulations that reduce the leakage directly for these operations (cf. approaches for the leakage reduction of gates for spin qubits \cite{wardrop2014,mehl2015-1} or superconducting qubits \cite{motzoi2009,rebentrost2009,egger2014}), there is also the need for an independent gate like operation, a ``leakage reduction unit'' (LRU) \cite{aliferis2007}. Many quantum error correction protocols only refocus qubit errors within the Hilbert space that codes the qubit, e.g., they refocus depolarizing or spin-flip errors \cite{preskill1998-2,nielsen2000}. It has been shown that if there is additionally a LRU, then fault-tolerant quantum computation can become tolerant to leakage errors \cite{aliferis2007}. 

A protocol to correct leakage errors, which is called ``leakage detection unit'' (LDU) in the following, was proposed by Gottesman \cite{gottesman1997} and Preskill \cite{preskill1998-2}. The authors described a gate sequence that detects leakage errors of a data qubit (called ``$\qu$'') by an ancilla qubit (called ``$\an$''). $\qu$ is used for the quantum computation, but $\an$ is only needed for the LDU. $\an$ is initialized to a known qubit state at the start of the LDU. The LDU inverts $\an$ if $\qu$ has not leaked, but $\an$ remains unchanged if leakage has occurred. Measuring the state of $\an$ determines if $\qu$ has leaked and $\qu$ needs to be reinitialized, or if $\qu$ is still a valid qubit state. $\an$ can be discarded after the LDU.

While the Gottesman-Preskill LDU uses measurements to detect if leakage has occurred, we will show that such a measurement process is indeed not necessary to correct for leakage using LRUs. We introduce two  generic approaches to construct LRUs, and apply them specifically to STQs. In every case, an ancilla qubit $\an$ is used as a resource to correct the leakage of the data qubit $\qu$. For the first LRU, $\an$ provides for $\qu$ a state from the computational subspace if leakage has occurred, but $\qu$ is untouched without leakage events. After this LRU, $\an$ can be discarded. For the second LRU, the state of $\qu$ is transferred to $\an$ only if there has been no leakage. For leakage events, $\qu$ keeps the leaked state, and $\an$ provides a new state from the computational subspace. In total, the definitions of $\qu$ and $\an$ are then interchanged after this LRU.

Recently, there were a few alternative studies of leakage and quantum error correction that treated specific error models of superconducting qubits \cite{fowler2013,ghosh2013}. These publications consider a specific entangling protocol, and the leakage to just one excited quantum state is taken into account. It was shown that a threshold for fault-tolerant quantum computation still exists in this specific scenario, even without additional LRUs \cite{fowler2013}. Note especially that the energetically excited leakage states have a natural decay rate to the computational subspace, where the standard quantum error correction protocols can be used. In contrast, leakage within the spin Hilbert space for spin-qubit encodings is more problematic because the qubit subspace does not necessarily contain the energetic ground state \cite{kempe2001}, and thermal relaxation drives a qubit {\it out of} the computational subspace.

Specifically for STQs, the data qubit $\qu$ can leak during the quantum computation from the computational basis with $s_z=0$ $\left\{\ket{S^\qu},\ket{T_0^\qu}\right\}$ to the $s_z=\pm1$ states $\ket{T_+^\qu}$ and $\ket{T_-^\qu}$. Note that for normal manipulations of STQs, a global external magnetic field is present and a leakage state is usually the energetic ground state of the Hilbert space. We identify this state with $\ket{T_+}$.
\footnote{
GaAs has a negative $g$ factor, but Si has a positive $g$ factor.\cite{winkler2010} Depending on the direction of the global magnetic field, also \protect{$\ket{T_-}$} can be the ground state.
}
For STQs, it is easy to initialize an ancilla qubit $\an$ to the singlet state $\ket{S^\an}$ \cite{hanson2007-2}. We describe LRUs that leave the state of $\qu$ unchanged if it has not leaked, but we swap the states of $\an$ and $\qu$ if leakage has occurred. A similar LRU was described\cite{kempe2001-2} and constructed \cite{fong2011} for the three-electron spin-qubit encoding. However, there has been no study of STQs, even though this qubit is suited to provide similar LRUs. We will show that STQs provide extremely short and efficient LRUs, ideally using as few as three calculation steps. These protocols can be implemented for arrays of DQDs in a setup that also realizes fault-tolerant quantum computation with the surface code at the cost of adding only few additional ancilla qubits to the edges of the surface code lattice.

The organization of this paper is as follows. \sref{sec:Implementation} describes the physical implementation of LRUs for STQs in an architecture suited for quantum error correction with the surface code. \sref{sec:Correction} describes LRUs using exchange interactions that are known to be well-controlled interactions between QDs in close proximity. \sref{sec:Summary} discusses the possibilities to use alternative interaction mechanism that are more suited for long-range couplings between QDs, and the findings of the paper are summarized.

\section{Practical Implementation of Leakage Reduction Units
\label{sec:Implementation}}
We introduce two approaches to realize LRUs for STQs. In every case one ancilla qubit is needed to correct for the leakage of a data qubit. A direct approach corrects the leakage of the data qubit, while the ancilla qubit is only needed during the process of the leakage correction. We call this operation {\it SWAP If Leaked} (SIL). Because one is usually working with a lattice of data and ancilla qubits, there is the freedom to interchange the definitions of data and ancilla qubits after the leakage correction step. We call this operation {\it SWAP If Not Leaked} (SINL). We will show that the surface code layout permits both leakage correction sequences; it is only that different gate sequences are needed to construct the leakage correction operations.

The SIL operation is defined similarly to earlier studies \cite{kempe2001-2,fong2011}. In the beginning $\an$ is initialized to $\ket{S^\an}$. The SIL operation leaves $\qu$ unchanged if it has not leaked, but it replaces $\qu$ with a state from the qubit Hilbert space if leakage has occurred. The truth table for the SIL for STQs is:
\begin{align}
\label{eq:SIL1}
\ket{S^\qu S^\an}
&\rightarrow\ket{S^\qu S^\an},\\
\label{eq:SIL2}
\ket{T_0^\qu S^\an}
&\rightarrow\ket{T_0^\qu S^\an},\\
\label{eq:SIL3}
\ket{T_+^\qu S^\an}
&\rightarrow\alpha_1\ket{S^\qu T_+^\an}+\beta_1\ket{T_0^\qu T_+^\an},\\
\ket{T_-^\qu S^\an}
&\rightarrow\alpha_2\ket{S^\qu T_-^\an}+\beta_2\ket{T_0^\qu T_-^\an}.
\label{eq:SIL4}
\end{align}
$\mathcal{A}$ and $\mathcal{D}$ indicate the logical function of the qubit, whereas the order of the states always corresponds to the positions of the physical qubits. The constants $\alpha_1$, $\alpha_2$, $\beta_1$, and $\beta_2$ are arbitrary. In general, a leaked state of $\qu$ cannot be reinitialized to the correct state before the leakage occurred because the point in time when leakage occurred is unknown, and a leakage state faces an uncontrolled phase evolution. In the next step, $\an$ is discarded and the ancilla can be used for a different calculation step.

We introduce also a modification of the SIL operation, where the positions of $\qu$ and $\an$ are interchanged after the leakage correction step. We call this operation SINL. The truth table for this LRU is:
\begin{align}
\label{eq:SINL1}
\ket{S^\qu S^\an}
&\rightarrow\ket{S^\an S^\qu},\\
\label{eq:SINL2}
\ket{T_0^\qu S^\an}
&\rightarrow\ket{S^\an T_0^\qu},\\
\label{eq:SINL3}
\ket{T_+^\qu S^\an}
&\rightarrow\alpha_1\ket{T_+^\an S^\qu}+\beta_1\ket{T_+^\an T_0^\qu},\\
\ket{T_-^\qu S^\an}
&\rightarrow\alpha_2\ket{T_-^\an S^\qu}+\beta_2\ket{T_-^\an T_0^\qu}.
\label{eq:SINL4}
\end{align}
The constants $\alpha_1$, $\alpha_2$, $\beta_1$, and $\beta_2$ are arbitrary again.

\begin{figure}[htb]
\centering
\includegraphics[width=.49\textwidth]{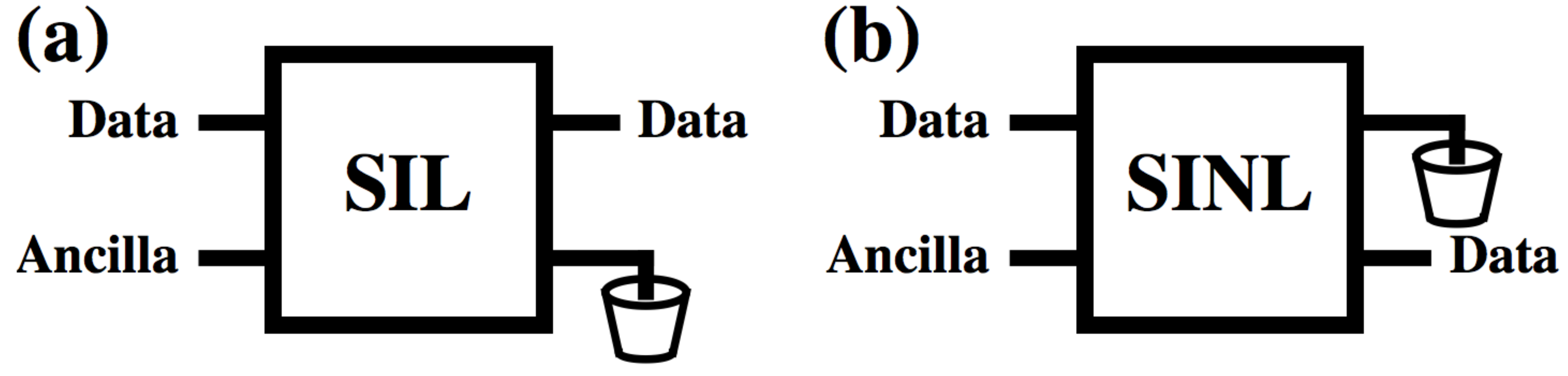}
\caption{Circuit diagrams for the {\it SWAP If Leaked} (SIL) and {\it SWAP If Not Leaked} (SINL) operations. The ancilla qubit need to be initialized to a known state at the beginning of the LRUs, and it can be discarded afterwards. (a) The positions of the data and ancilla qubits remain formally unchanged after the SIL operation. (b) The SINL operation interchanges the positions of the data and ancilla qubits.
\label{fig:01}}
\end{figure}

\fref{fig:01} shows the circuit diagrams of the SIL and SINL operations. For the SIL operation in \fref{fig:01}(a), the position of the data qubit remains unchanged. For the SINL operation in \fref{fig:01}(b), the positions of the data and ancilla qubit are interchanged after the leakage correction step.

Standard quantum error correction protocols neglect leakage errors out of the computational subspace. The surface code is one of the most prominent quantum error correction codes \cite{raussendorf2007,fowler2009,fowler2012,horsman2012,terhal2013}, and tolerates errors of the gate operations, the qubit initializations, and the readout of every qubit below a threshold of about $1\%$ error per operation. This protocol is especially promising because the error corrections and the manipulations of the encoded quantum information only requires nearest-neighbor interactions between neighboring physical qubits on a lattice.

\begin{figure}[htb]
\centering
\includegraphics[width=.4\textwidth]{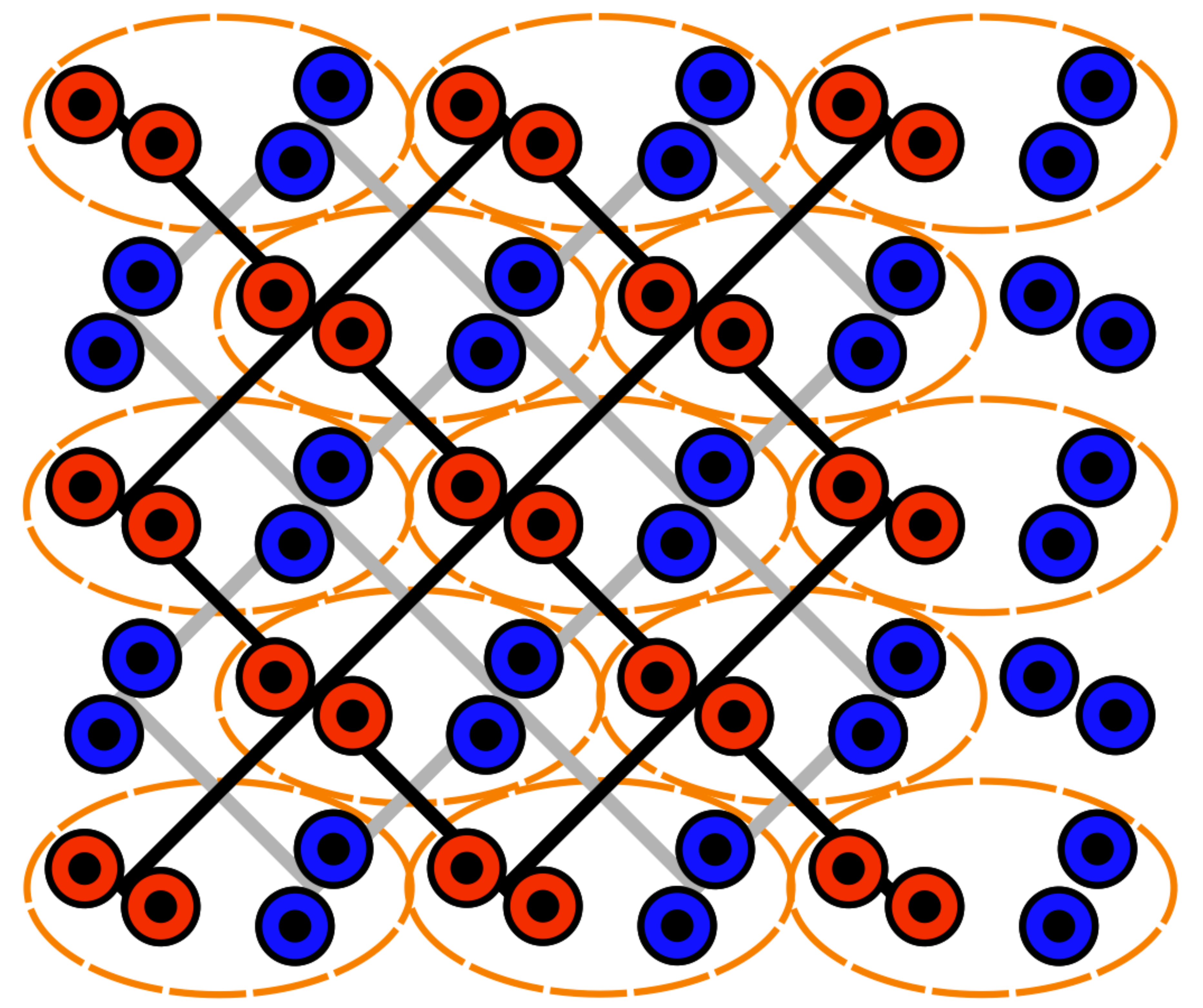}
\caption{Setup for fault-tolerant quantum computation with the surface code that also tolerates small leakage errors. A two-electron DQD encodes a qubit. The red DQDs are the data qubits, the blue DQDs are the ancilla qubits. The data qubits lie on the vertices of the black rectangular lattice, the ancilla qubits lie on the vertices of the gray lattice that is the dual lattice of the black one. Note that these lattices do not represent any physical interactions, and only nearest-neighbor interactions between a data qubit with each of the surrounding ancilla qubits are needed. Furthest to the right are some additional ancilla qubits that are needed for the LRUs. The ancilla qubits can serve both as the syndrome qubits for the quantum error correction in the surface code and as the ancilla qubits for the leakage corrections. In the LRUs each data qubit must interact with one ancilla qubit, as sketched by the orange circles around pairs of neighboring DQDs.
\label{fig:02}}
\end{figure}

The surface code should be reviewed briefly. One error-corrected qubit is stored in a rectangular lattice of $n\times n$ physical qubits, as sketched in \fref{fig:02}. The red DQDs are the data qubits that encode the quantum information, the blue DQDs are the ancilla qubits that are only needed for the quantum error corrections and the manipulation of the quantum information. Specifically in the case of STQs, one physical qubit is always coded using a two-electron DQD. The lattices of the data qubits and the ancilla qubits are shifted relative to each other, and each data qubit is surrounded only by ancilla qubits. The ancilla qubits are used to measure the parity of the wave function of the surrounding data qubits, which is sufficient to detect qubit errors within the computational subspace.

The surface code setup can be used without changes for the LRUs of \fref{fig:01}. Because the ancilla qubits do not store any relevant information after the parity check operations of the surface code, these qubits can be initialized to a singlet state and the leakage correction procedure can be executed. One ancilla qubit is needed for every data qubit. In the setup of \fref{fig:02}, we therefore add additional ancilla qubits to the edges of the surface code lattice. For the SINL operation, the definitions of data and ancilla qubits swap after the leakage correction procedure, which results in a shift of the surface code layout after one leakage correction step. Note that two similar approaches to include leakage corrections to quantum error correction protocols were discussed recently \cite{ghosh2014,suchara2014}.

The realization of our LRU does not cause a large overhead in the number of ancilla qubits that are required for an encoded qubit. The number of ancilla qubits only increases linearly in the size of the surface by $\mathcal{O}\left(n\right)$, while each error-corrected qubit needs $\mathcal{O}\left(n^2\right)$ qubits. A study of error-corrected qubits that are sufficient for quantum computation suggests patch sizes of $n>60$ \cite{jones2012}. Also the addition of LRUs to the surface code algorithm does not increase the number of gate operations significantly. One round of leakage correction requires one additional LRU for every data qubit. In contrast, one data qubit is involved in four parity check operations in one round of the surface code error correction.

\section{Leakage Correction Sequences
\label{sec:Correction}}
\begin{figure}[htb]
\centering
\includegraphics[width=.49\textwidth]{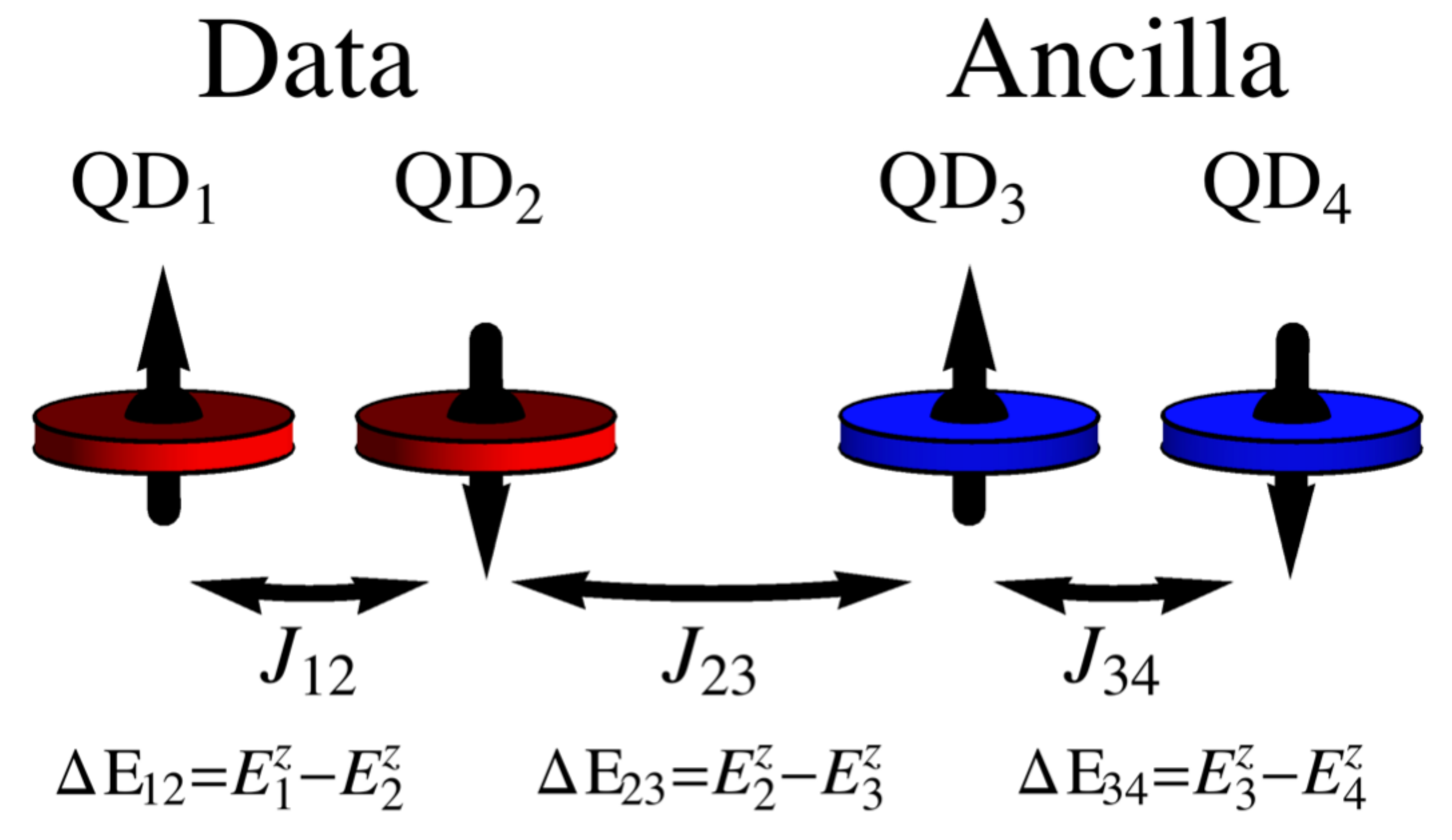}
\caption{Setup for the LRU of a data qubit (red DQD) with an ancilla qubit (blue DQD). The electron tunnelings between neighboring QDs are allowed, giving the exchange interactions $J_{12}$, $J_{23}$, and $J_{34}$. The magnetic fields at each QD can be independently prepared, defining the magnetic field gradients $\Delta E_{12}$, $\Delta E_{23}$, and $\Delta E_{34}$.
\label{fig:03}}
\end{figure}

We specifically describe LRUs for one data qubit $\qu$ and one ancilla qubit $\an$, as sketched in \fref{fig:03}. We consider a setup of DQDs in close proximity, where the electron transfer between the DQDs is possible, such that the DQDs are coupled by exchange interaction \cite{levy2002}. It does not matter if there is a direct exchange interaction between the DQDs, or if the exchange interaction is mediated via another QD as in \rcite{mehl2014-1}. We will discuss the possibility to use long-range interactions for the leakage correction of STQs in \sref{sec:Summary}.

Universal single-qubit control of $\qu$, which is coded using QD$_1$ and QD$_2$, is provided using the exchange interaction $\frac{J_{12}}{4} \left(\bm{\sigma}_1\cdot\bm{\sigma}_2-\bm{1}\right)$ and a magnetic field gradient $\frac{\Delta E_{12}}{2}\left(\sigma_1^z-\sigma_2^z\right)$. $\bm{\sigma}_i=\left(\sigma^x_i,\sigma^y_i,\sigma^z_i\right)^T$ is the vector of Pauli operators for the electrons at QD$_i$, and $\Delta E_{12}=E_{1}^z-E_{2}^z$ is the energy difference arising from the local magnetic fields at QD$_1$ and QD$_2$. We assume we have independent control over $J_{12}$ and $\Delta E_{12}$. In reality, more complicated manipulation protocols will likely be needed and the approach of \rcite{cerfontaine2014} can be applied to the following gates if the need arises. A single-qubit phase gate of $\qu$ is described by the time evolution $Z_\phi^\qu=e^{-i\left[2\pi\frac{\phi}{2}\left(\sigma^z_1-\sigma^z_2\right)\right]}$, where the rotation angle is determined by $\phi=\frac{\Delta E_{12}t}{h}$.\footnote{
Our notation refers to a Bloch sphere\cite{nielsen2000} where \protect{$\ket{\uparrow\downarrow}$} and \protect{$\ket{\downarrow\uparrow}$} lie on the poles, and \protect{$\ket{S}$} and \protect{$\ket{T_0}$} are on the equator. $Z_\phi$ ($X_\phi$) describes a rotation around the z axis (x axis).
}
A similar labeling is used for exchange gates with $X_\phi^\qu=e^{-i\left[2\pi\frac{\phi}{4}\left(\bm{\sigma}_1\cdot\bm{\sigma}_2-\bm{1}\right)\right]}$, for $\phi=\frac{J_{12}t}{h}$.\cite{Note2} The equivalent definitions are used for $\an$, which is coded using QD$_3$ and QD$_4$, giving the phase gate $Z_\phi$ and the exchange gate $X_\phi$.

The LRU additionally requires an interaction between $\qu$ and $\an$; we assume that the exchange operation $J_{23}$ between QD$_2$ and QD$_3$ can be controlled by electric gates. It can be desirable that the magnetic fields at QD$_2$ and QD$_3$ differ during these entangling operations, giving the energy difference $\Delta E_{23}=E_{2}^z-E_{3}^z$ and the entangling gate:
\begin{align}
\mathcal{U}_{\phi,\psi}=e^{-i\left\{
2\pi
\left[
\frac{\phi}{4}\left(\bm{\sigma}_2\cdot\bm{\sigma}_3-\bm{1}\right)+
\frac{\psi}{2}\left(\sigma_2^z-\sigma_3^z\right)
\right]
\right\}},
\label{eq:Entangle}
\end{align}
for $\phi=\frac{J_{23}t}{h}$ and $\psi=\frac{\Delta E_{23}t}{h}$. Note that setting the magnetic fields at QD$_1$ and QD$_4$ to the same value is not strictly required; differences of the magnetic fields at QD$_1$ and QD$_4$ can be corrected by single-qubit gates of $\qu$ and $\an$.

In general, the universal control of the spin Hilbert space of four electrons also requires the control over the spatially homogeneous global magnetic field and the relative magnetic fields between the DQD pairs. For our gate sequences, these magnetic fields are not specified because they are not explicitly needed in our gate constructions. Note especially that a spatially homogeneous global magnetic field across the four QDs is still desirable for STQs because it reduces the leakage from the computational subspace. Such a magnetic field is irrelevant for our LRUs because it commutes with all the described interactions. It will provide in total only an overall phase factor between the different total $s_z$ subspaces.

\begin{figure}[htb]
\centering
\includegraphics[width=.49\textwidth]{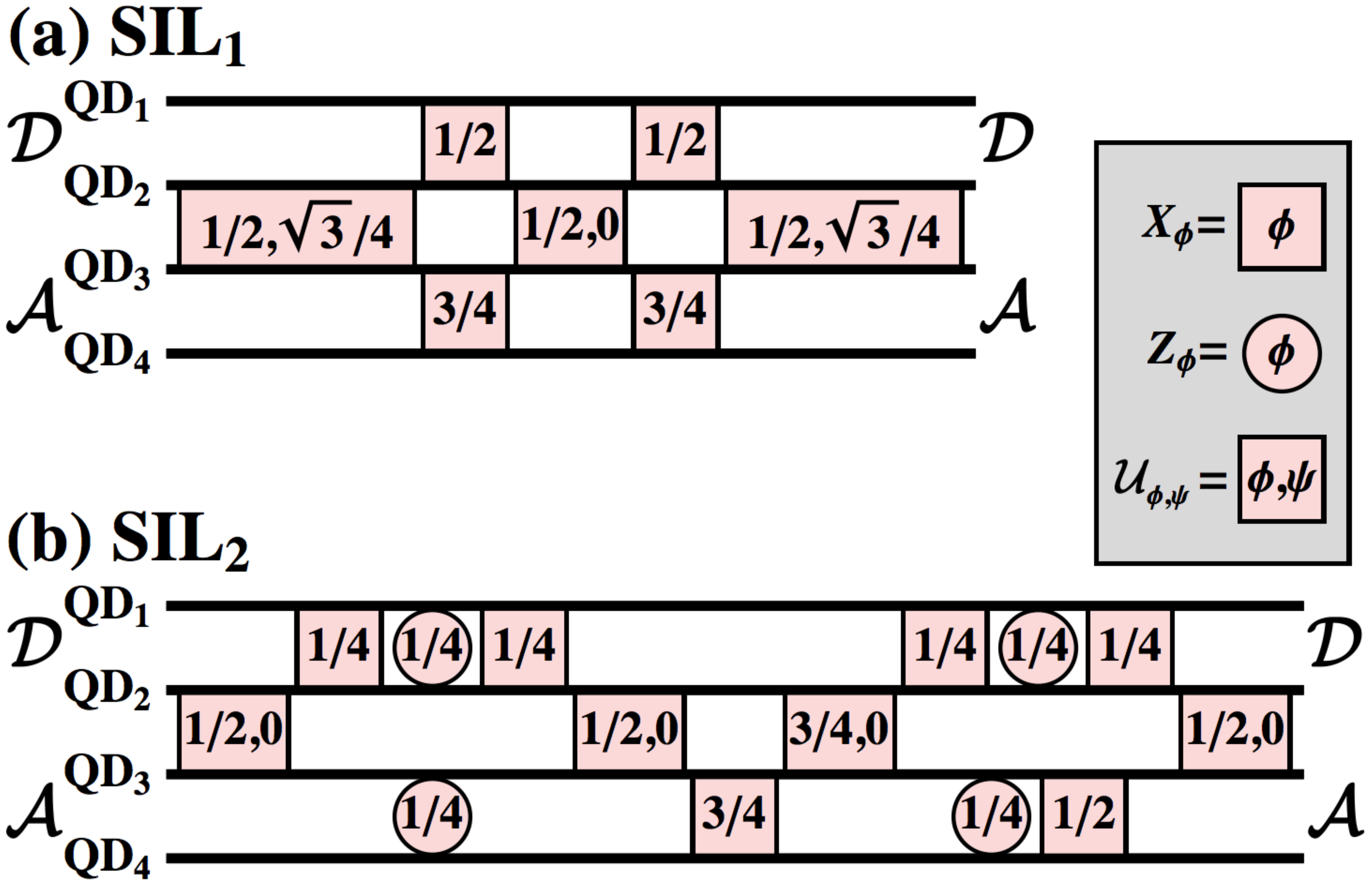}
\caption{
Gate operations for the SIL operation, according to \meref{eq:SIL1}-\eqref{eq:SIL4}. $Z_\phi$ and $X_\phi$ are the phase and exchange gates, and $\mathcal{U}_{\phi,\psi}$ is the effective interaction between the electrons at QD$_2$ and QD$_3$ according to \eref{eq:Entangle}. (a) If the magnetic fields at QD$_2$ and QD$_3$ differ, SIL is constructed in a five step gate sequence. (b) If the magnetic fields at QD$_3$ and QD$_4$ are identical, then eleven gate operations are needed.
\label{fig:04}}
\end{figure}

We use a numerical search algorithm to find the SIL gate sequence according to \meref{eq:SIL1}-\eqref{eq:SIL4}, similarly to earlier studies \cite{divincenzo2000,fong2011} (cf. \rcite{mehl2014-1} for a more detailed description of the search algorithm). \fref{fig:04}(a) is the simplest SIL sequence we found, which we call SIL$_1$, when a magnetic field gradient between QD$_2$ and QD$_3$ is present. SIL$_1$ needs three interactions between $\qu$ and $\an$. $\mathcal{U}_{\frac{1}{2},\frac{\sqrt{3}}{4}}$ is the same gate that we used in \rcite{mehl2014-1} to entangle two STQs. $\mathcal{U}_{\frac{1}{2},0}$ is a $\text{SWAP}$ operation between the spins at QD$_2$ and QD$_3$. The constants of \meref{eq:SIL3}-\eqref{eq:SIL4} are $\alpha_1=\alpha_2=\beta_2=e^{i\frac{3\pi}{4}}/\sqrt{2}$ and $\beta_1=e^{-i\frac{\pi}{4}}/\sqrt{2}$. For $\mathcal{U}_{\frac{1}{2},0}$, an evolution only under $J_{23}$ is needed, while $\Delta E_{23}$ must be turned to zero. It might be favorable to replace $\mathcal{U}_{\frac{1}{2},0}$ by $\mathcal{U}_{\frac{1}{2\sqrt{2}},\frac{1}{4\sqrt{2}}}\mathcal{U}_{0,\frac{1}{4}}\mathcal{U}_{\frac{1}{2\sqrt{2}},\frac{1}{4\sqrt{2}}}$ because, in general, fast modifications of $\Delta E_{23}$ are difficult \cite{hanson2007-1}.

If there is no magnetic field gradient between QD$_2$ and QD$_3$, then we need four exchange operations between $\qu$ and $\an$ for the SIL operation. This gate, which we call SIL$_2$, is shown in \fref{fig:04}(b). The single-qubit gate operations are just a specific choice, and they can be substituted by other gate sequences. The constants according to \meref{eq:SIL3}-\eqref{eq:SIL4} are $\alpha_1=\beta_1=\alpha_2=e^{i\frac{3\pi}{4}}/\sqrt{2}$ and $\beta_2=e^{-i\frac{\pi}{4}}/\sqrt{2}$.

\begin{figure}[htb]
\centering
\includegraphics[width=.49\textwidth]{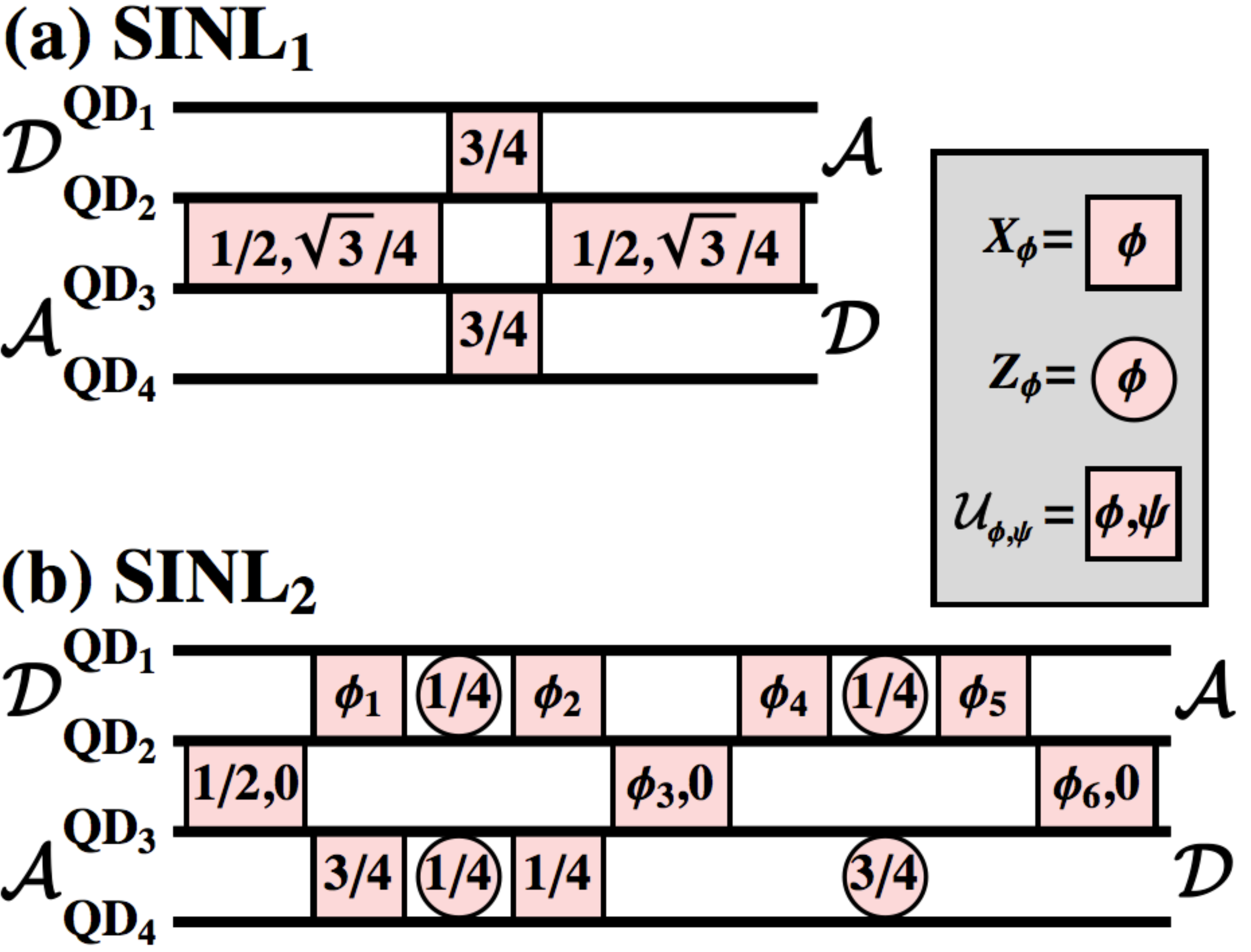}
\caption{
Gate operations for the SINL operation, according to \meref{eq:SINL1}-\eqref{eq:SINL4}. (a) If the magnetic fields at QD$_2$ and QD$_3$ differ, then three gate sequences are needed for the SINL. (b) For identical magnetic fields at QD$_2$ and QD$_3$ nine gate operations are needed. The parameters for $\phi_1$-$\phi_6$ are given in \aref{app:Cons}.
\label{fig:05}}
\end{figure}

\fref{fig:05} shows the SINL sequences according to \meref{eq:SINL1}-\eqref{eq:SINL4}. Note that the designations of $\qu$ and $\an$ reverse after these gate sequences. If the magnetic fields at QD$_2$ and QD$_3$ can differ, then only three operations are needed to construct the SINL operation. \fref{fig:05}(a) shows this gate sequence, which we call SINL$_1$. The constants of \meref{eq:SINL3}-\eqref{eq:SINL4} are $\alpha_1=\alpha_2=\beta_2=e^{i\frac{\pi}{4}}/\sqrt{2}$ and $\beta_1=e^{-i\frac{3\pi}{4}}/\sqrt{2}$. We also found a SINL operation for identical magnetic field at QD$_2$ and QD$_3$ in a nine step sequence. The constants of SINL$_2$ in \fref{fig:05}(b) are given in \aref{app:Cons}.

\section{Discussion and Conclusion
\label{sec:Summary}}
In principle, it is possible to use long-range interactions to construct LRUs for STQs instead of the short-range Heisenberg interactions. Coulomb interactions\cite{taylor2005,trifunovic2012,srinivasa2014} or cavity-mediated couplings\cite{burkard2006,taylor2006-2} between STQs have been suggested to couple distant STQs. Both coupling mechanisms can be described by an effective two-qubit interaction $\mathcal{I}\tau_z^\qu\tau_z^\an$, which acts only on the qubit subspace, with an effective coupling constant $\mathcal{I}$ and $\tau_z=\op{T_0}{T_0}-\op{S}{S}$. One can construct the SINL operation according to
\begin{align}
H^\qu
e^{-i\frac{\pi}{4}\tau_z^\qu\tau_z^\an}
e^{-i\frac{3\pi}{4}\tau_x^\qu}
e^{-i\frac{3\pi}{4}\tau_x^\an}
e^{-i\frac{\pi}{4}\tau_z^\qu\tau_z^\an}
H^\an,
\label{eq:iSWAP}
\end{align}
with $\tau_x=\op{S}{T_0}+\op{S}{T_0}$. $H^{\qu}$ and $H^{\an}$ are the Hadamard gates for $\qu$ and $\an$. The SIL operation cannot be realized with the $\mathcal{I}\tau_z^\qu\tau_z^\an$ interaction because the $s_z$ quantum number remains unchanged at each QD. Even though the first attempts to realize Coulomb\cite{weperen2011,shulman2012} and cavity-mediated\cite{frey2012,toida2013} two-qubit operations have been made, it still remains an open problem to raise the effective interaction strength $\mathcal{I}$ to sufficiently high magnitudes that allow high-fidelity entangling operations before the qubit dephases.

Our study has shown that an array of DQDs realizes a setup for fault-tolerant quantum computation of STQs that even tolerates leakage errors. Earlier studies have shown that high-fidelity single-qubit \cite{cerfontaine2014} and two-qubit \cite{taylor2005,mehl2014-1,srinivasa2014} gate operations can indeed be realized theoretically in these systems. Experiments have realized excellent single-qubit gates \cite{petta2005,foletti2009,bluhm2010,shulman2014,maune2012,wu2014}, while high-fidelity two-qubit gates are still to be done. The initialization and the readout of STQs can be done with high fidelities, such that fault-tolerant quantum computation can readily be implemented. To additionally include LRUs, we proposed a lattice of DQDs, where the exchange operations between QDs of the data qubit and the ancilla qubit can be controlled. In the ideal setup, with different magnetic fields at the involved, neighboring QDs, our LRU only requires three calculation steps.

Our described LRUs use one ancilla qubit for every coded qubit, while the ancilla qubits are only needed during the leakage corrections. We describe two methods for leakage corrections. In one case, the ancilla qubits are only used as a resource to provide a state from the qubit subspace if leakage has occurred. In the other case, the data qubit and the ancilla qubit change their positions if no leakage has occurred. Because the ancilla qubits are required anyway in standard quantum error correction protocols, both approaches to construct LRUs are equally permitted. We find that the freedom of moving the quantum information by one lattice site during the leakage correction step generally results in shorter gate sequences.

None of our LRUs require any measurements opposed to LDUs that were proposed earlier \cite{gottesman1997,preskill1998-2}. LRUs are especially superior over LDUs if the measurement process is time consuming or disturbed by errors. Our LRUs can easily be added to the surface code error correction codes to achieve fault-tolerant quantum computation. They will neither add a large overhead in the number of required ancilla qubits, nor in the calculation steps compared to standard error correction codes.

We briefly discuss the possibility of leakage propagation in surface code calculations of STQs. Our LRUs map a leakage error to a regular gate error. One should notice that an uncorrected leaked qubit does not introduce additional leakage during the regular error-correction step. Exchange interactions leave the $s_z$ quantum numbers of the involved spins unchanged, which only provides the possibility to transfer leakage, but exchange gates cannot produce additional leakage events. Coulomb interactions and cavity-mediated couplings cannot transport leaked qubits, and therefore they can only cause a propagation of qubit errors. In total for quantum error correction protocols with STQs, leaked qubits catalyze gate errors, but two-qubit gates between STQs cannot create additional leaked qubits from an uncorrected leaked qubit. A thermal accumulation of leakage states for STQs will ultimately destroy the possibility to achieve fault tolerant quantum computation. If we add LRUs, we are able to stabilize the leakage at the level of a single LRU. Altogether we only increase the effective gate errors in the surface code and expect that the threshold criterion of the surface code should persist when we add LRUs.

Our study can be continued with an in-depth analysis of specific error models for spin qubits to describe leakage errors in addition to the usual gate, initialization, and readout errors. It is especially important to analyze the consequences of imperfect leakage correction sequences more quantitatively. Decoherence is the main obstacle to construct high-fidelity quantum gates for spin qubits (cf., e.g., \mrcite{hanson2007-2}{zwanenburg2013}), and it will also disturb our leakage correction protocols such that leakage errors are only partly recovered.
Furthermore our study should bring attention to the problem of leakage errors in the field of quantum computation with spin qubits, where it has received little attention so far. Not only can our proposed LRUs mitigate leakage errors, but they also show that leakage errors do not present a fundamental problem for fault-tolerant quantum computation.

\begin{acknowledgements}
S.M. and D.P.D. are grateful for support from the Alexander von Humboldt foundation.
\end{acknowledgements}

\appendix
\section{Numerical Values for SINL$_2$
\label{app:Cons}}
We give the numerical values for the constants $\phi_1-\phi_6$ of the gate SINL$_2$ of \fref{fig:05}(b).  Independent numerical simulations gave four sets of equivalent gate operations, called $\left(1\right)-\left(4\right)$, with the numerical values:

\begin{align}
\phi_1^{\left(1\right)} = 0.345073936796977,\\
\phi_2^{\left(1\right)} = 0.130451628557808,\\
\phi_3^{\left(1\right)} = 0.391184696119253,\\
\phi_4^{\left(1\right)} = 0.854636869769667,\\
\phi_5^{\left(1\right)} = 0.676562387880084,\\
\phi_6^{\left(1\right)} = 0.687295455441529,
\end{align}

\begin{align}
\phi_1^{\left(2\right)} = 0.154926063203023,\\
\phi_2^{\left(2\right)} = 0.369548371442192,\\
\phi_3^{\left(2\right)} = 0.608815303880747,\\
\phi_4^{\left(2\right)} = 0.145363130230333,\\
\phi_5^{\left(2\right)} = 0.323437612119916,\\
\phi_6^{\left(2\right)} = 0.312704544558471,
\end{align}

\begin{align}
\phi_1^{\left(3\right)} = 0.351157090810363,\\
\phi_2^{\left(3\right)} = 0.929368971476208,\\
\phi_3^{\left(3\right)} = 0.220608581536442,\\
\phi_4^{\left(3\right)} = 0.584927407435767,\\
\phi_5^{\left(3\right)} = 0.298820462202286,\\
\phi_6^{\left(3\right)} = 0.340060072262521,
\end{align}

\begin{align}
\phi_1^{\left(4\right)} = 0.148842909189637,\\
\phi_2^{\left(4\right)} = 0.570631028523793,\\
\phi_3^{\left(4\right)} = 0.779391418463558,\\
\phi_4^{\left(4\right)} = 0.415072592564233,\\
\phi_5^{\left(4\right)} = 0.701179537797713,\\
\phi_6^{\left(4\right)} = 0.659939927737479.
\end{align}

\bibliography{library}
\end{document}